\documentclass[a4paper,12pt]{article}

\usepackage[latin1]{inputenc}
\usepackage[T1]{fontenc}
\usepackage{graphicx}
\usepackage{amsfonts}
\usepackage{amssymb}
\usepackage{amsmath} 
\usepackage{amsthm}
\usepackage{url}

\usepackage{natbib}

\usepackage[margin=1.2in]{geometry}

\begin{document}

\title{Epitope prediction improved by multitask support vector
  machines}
\author{
Laurent Jacob\\
Centre for Computational Biology\\
Ecole des Mines de Paris\\
35, rue Saint-Honor\'e\\
77300 Fontainebleau, France\\
\texttt{laurent.jacob@ensmp.fr}
\and
Jean-Philippe Vert\\
Centre for Computational Biology\\
Ecole des Mines de Paris\\
35, rue Saint-Honor\'e\\
77300 Fontainebleau, France\\
\texttt{jean-philippe.vert@ensmp.fr}
}

\maketitle

\begin{abstract}
    \emph{In silico} methods for the prediction of antigenic peptides
    binding to MHC class I molecules play an increasingly important role
    in the identification of T-cell epitopes. Statistical and machine
    learning methods, in particular, are widely used to score candidate
    epitopes based on their similarity with known epitopes and non
    epitopes. The genes coding for the MHC molecules, however, are highly
    polymorphic, and statistical methods have difficulties to build models
    for alleles with few known epitopes. In this case, recent works have
    demonstrated the utility of leveraging information across alleles to
    improve the performance of the prediction.

    We design a support vector machine algorithm that is able to learn
    epitope models for all alleles simultaneously, by sharing information across
    similar alleles. The sharing of information across alleles is
    controlled by a user-defined measure of similarity between alleles. We
    show that this similarity can be defined in terms of supertypes, or
    more directly by comparing key residues known to play a role in the
    peptide-MHC binding. We illustrate the potential of this approach on various
    benchmark experiments where it outperforms other state-of-the-art
    methods.
\end{abstract}

\section{Introduction}

A key step in the immune response to pathogen invasion is the
activation of cytotoxic T-cells, which is triggered by the recognition
of a short peptide, called epitope, bound to Major Histocompatibility
Complex (MHC) class I molecules and presented to the T-cells. This
recognition is supposed to trigger cloning and activation of cytotoxic
lymphocytes able to identify and destroy the pathogen or infected
cells. MHC class I epitopes are therefore potential tools for the
development of peptide vaccines, in particular for AIDS
vaccines~\citep{McMichael2002quest}. They are also potential tools for
diagnosis and treatment of cancer~\citep{Wang1999Human,Sette2001HLA}.

Identifying MHC class I epitope in a pathogen genome is therefore
crucial for vaccine design. However, not all peptides of a pathogen
can bind to the MHC molecule to be presented to T-cells: it is
estimated that only 1 in 100 or 200 peptides actually binds to a
particular MHC~\citep{Yewdell1999Immunodominance}. In order to
alleviate the cost and time required to identify epitopes
experimentally, \emph{in silico} computational methods for epitope
prediction are therefore increasingly used. Structural approaches, on
the one hand, try to evaluate how well a candidate epitope fit in the
binding groove of a MHC molecule, by various threading or docking
approaches~\citep{Rosenfeld1995Flexible,Schueler-Furman2000Structure-based,Tong2006Prediction,Bui2006Structural}.
Sequence-based approaches, on the other hand, estimate predictive
models for epitopes by analyzing and learning from sets of known
epitopes and non-epitopes. Models can be based on
motifs~\citep{Rotzschke1992Peptide, Rammensee1995MHC}, profiles
\citep{Parker1994Scheme,Rammensee1999SYFPEITHI,Reche2002Prediction},
or machine learning methods like artificial neural
networks~\citep{Honeyman1998Neural,Milik1998Application,Brusic2002Prediction,Buus2003Sensitive,
  Nielsen2003Reliable,Zhang2005MULTIPRED}, hidden Markov
models~\citep{Mamitsuka1998Predicting}, support vector machines
(SVM)~\citep{Donnes2002Prediction,Zhao2003Applicationa,Bhasin2004Prediction,Salomon2006Predicting},
boosted metric learning~\citep{Hertz2006PepDist} or logistic
regression~\citep{Heckerman2006Leveraging}. Finally, some authors have
recently proposed to combine structural and sequence-based approaches~\citep{Antes2006DynaPred,Jojic2006Learning}. Although
comparison is difficult, sequence-based approaches that learn a model
from the analysis of known epitopes benefit from the accumulation of
experimentally validated epitopes and will certainly continue to
improve as more data become available.

The binding affinity of a peptide depends on the MHC molecule's 3D
structure and physicochemical properties, which in turns vary between
MHC alleles. This compels any prediction method to be allele-specific:
indeed, the fact that a peptide can bind to an allele is neither
sufficient nor necessary for it to bind to another allele. Since MHC
genes are highly polymorphic, little training data if any is available
for some alleles. Thus, though achieving good precisions in general,
classical statistical and machine learning-based MHC-peptide binding
prediction methods fail to efficiently predict bindings for these
alleles.

Some alleles, however, can share binding properties. In particular,
experimental work~\citep{Sidney1995Several, Sidney1996Definition,
  Sette1998HLA, Sette1999Nine} shows that different alleles have
overlapping peptide repertoires. This fact, together with the
posterior observation of structural similarities among the alleles
sharing their repertoires allowed the definition of HLA allele
supertypes, which are families of alleles exhibiting the same behavior
in terms of peptide binding. This suggests that sharing information
about known epitopes across different but similar alleles has the
potential to improve predictive models by increasing the quantity of
data used to establish the model. For
example,~\citet{Zhu2006Improving} show that simply pooling together
known epitopes for different alleles of a given supertype to train a
model can improve the accuracy of the model. \citet{Hertz2006PepDist}
pool together epitope data for all alleles simultaneously to learn a
metric between peptides, which is then used to build predictive models
for each allele. Finally, \citet{Heckerman2006Leveraging} show that
leveraging the information across MHC alleles and supertypes
considerably improves individual allele prediction accuracy.

In this paper we show how this strategy of leveraging information
across different alleles when learning allele-specific epitope
prediction models can be naturally performed in the context of SVM, a
state-of-the-art machine learning algorithm. This new formulation is
based on the notion of \emph{multitask
  kernels}~\citep{Evgeniou2005Learning}, a general framework for
solving several related machine learning problems simultaneously.
Known epitopes for a given allele contribute to the model estimation
for all other alleles, with a weight that depends on the similarity
between alleles.  Here the notion of similarity between alleles can be
very general; we can for example
follow~\citet{Heckerman2006Leveraging} and define two alleles to be
similar if they belong to the same supertype, but the flexibility of
our mathematical formulation also allows for more subtle notions of
similarity, based for example of sequence similarity between alleles.
On a benchmark experiment we demonstrate the relevance of the
multitask SVM approach which outperforms state-of-the-art prediction
methods.

\section{Methods}

In this section, we explain how information can be shared between
alleles when SVM models are trained on different alleles. For the sake
of clarity we first explain the approach in the case of linear
classifiers, and then generalize it to more general models.

\subsection{Sharing information with linear classifiers}
Let us first assume that epitopes are represented by $d$-dimensional
vectors $x$, and that for each allele $a$ we want to learn a linear
function $f_{a}(x) = w^\top x$ to discriminate between epitopes
and non-epitopes, where $w\in\mathbb{R}^d$. A natural way to share information between different
alleles is to assume that each vector $w$ is the sum of a common
vector $w_{c}$ which is common to all alleles, and of an
allele-specific vector $w_{a}$, resulting in a classifier:
\begin{equation}
  \label{eq:mtf}
f_{a}(x) = (w_{c}+w_{a})^\top x\,. 
\end{equation}
In this equation the first term $w_{c}$ accounts for general
characteristics of epitopes valid for all alleles, while the second
term $w_{a}$ accounts for allele-specific properties of epitopes. In
order to estimate such a model from data it is convenient to rewrite
it as a simple linear model in a larger space as follows. Assuming
that there are $p$ alleles $\left\{a_{1},\ldots,a_{p}\right\}$ we can
indeed rewrite \eqref{eq:mtf} as:
\begin{equation}
  \label{eq:mtfbis}
f_{a}(x) = W^\top \Phi(a,x)\,,
\end{equation}
where $W$ is the $d\times(p+1)$-dimensional vector $W =
\left(w_{c}^\top,w_{a_{1}}^\top,\ldots,w_{a_{p}}^\top\right)^\top$ and
$\Phi(a,x)=
(x^\top,0^\top,\ldots,0^\top,x^\top,0^\top,\ldots,0^\top)^\top \in
\mathbb{R}^{d\times(p+1)}$ is the vector obtained by concatenating the
vector $x$ with $p$ blocks of zeros, except for the $a$-th block which
is a copy of $x$. Indeed it is then easy to check that $W^\top
\Phi(a,x) = (w_{c}+w_{a})^\top x$, hence that \eqref{eq:mtf} and
\eqref{eq:mtfbis} are equivalent. Each (peptide,allele) pair is therefore
mapped to a large vector $\Phi(x,a)$ with only two non-zero parts, one
common to all alleles and one at an allele-specific position.

The parameters of this model, namely the weights $w_{c}$ and $w_{a}$
for all alleles $a$, can then be learned simultaneously by any linear
model, such as logistic regression or SVM, that estimates a vector $W$ in \eqref{eq:mtfbis} from a training set
$\left((x_{1},a_{1},y_{1}),\ldots,(x_{n},a_{n},y_{n})\right)$ of
(peptide,allele) pairs labeled as $y_{i}=+1$ if peptide $x_{i}$ is an
epitope of allele $a_{i}$, $y_{i}=-1$ otherwise. This approach was
followed by~\citet{Heckerman2006Leveraging} who included another level
of granularity to describe how information is shared across alleles,
by considering allele-specific, supertype-specific and common weight
vectors.

In summary, it is possible to embed the allele information in the
description of the data point to estimate linear models in the new
$\text{peptide} \times \text{allele}$ space to share information
across alleles. It is furthermore possible to adjust how information
is shared by choosing adequate functions $\Phi(x,a)$ to represent
(peptide,allele) pairs. In other words, it is possible to consider the
problem of leveraging across the alleles as a simple choice of
representation, or feature design for the (peptide,allele) pairs that
are to be used to learn the classifier. This approach, however, is
limited by at least two constraints:
\begin{itemize}
\item It can be uneasy to figure out how to represent the allele
  information in the mapping $\Phi(x,a)$.
  In~\citealp{Heckerman2006Leveraging}, this is done \emph{via}
  Boolean conjunctions and leads to a convenient form for the
  prediction functions, like \eqref{eq:mtf} with a third term
  accounting for the supertype. Including more prior knowledge
  regarding when two alleles should share more information, \emph{e.g.},
  based on structural similarity between alleles, is however not an
  easy task.
  \item Practically, injecting new features in the vector $\Phi(x,a)$ increases the dimension of
  the space, making statistical estimation, storage, manipulation and optimization tasks much harder.
\end{itemize}
In the next subsection we show how both limitations can be overcome by reformulating this approach in the framework of kernel methods.

\subsection{The kernel point of view}
SVM, and more generally kernel methods, only access data through the
computation of inner products between pairs of data points, called a
\emph{kernel} function~\citep{Vapnik1998Statistical,Scholkopf2002Learning,Schoelkopf2004Kernel}.
As a result, estimating the weights $W$ in (\ref{eq:mtfbis}) with a
SVM does not require to explicitly compute or store the vectors
$\Phi(x,a)$ for training and test pairs of alleles and peptides.
Instead, it only requires to be able to compute the kernel between any
two pairs $(x,a)$ and $(x',a')$ given, in our linear example, by:
\begin{align*}
  K\left((x,a),(x',a')\right) =& \Phi(x,a)^\top \Phi(x',a')\\
  =& \begin{cases}
  2 x^Tx'&\text{ if }a=a'\,,\\
  x^Tx'&\text{ if }a\neq a'\,.\\
  \end{cases}
\end{align*}
Let us now introduce the following two kernels, respectively between peptides only and between alleles only:
\begin{align*}
  & K_{pep}(x,x') \stackrel{\Delta}{=} x^\top x'\\
  & K_{all}(a,a') \stackrel{\Delta}{=} \begin{cases}
  2 &\text{ if }a=a'\,,\\
  1&\text{ if }a\neq a'\,.\\
  \end{cases}
\end{align*}
It is easy to see that both kernels are valid positive definite kernels for peptides and alleles, respectively.
With these notations we see that the kernel for pairs $(x,a)$ can be expressed as the \emph{product} of the kernel for alleles and the kernel for peptides:
\begin{equation}
\label{eq:product}
K\left((x,a),(x',a')\right) = K_{all}(a,a')K_{pep}(x,x') \, ,
\end{equation}
which is also the kernel associated to the tensor product space of the
Hilbert spaces associated to $K_{pep}$ and $K_{all}$~\citep{Aronszajn1950Theory}. Such kernels are
used in particular in the field of \emph{multitask
  learning}~\citet{Evgeniou2005Learning}, where several related
machine learning tasks must be solved simultaneously. The allele
kernel $K_{all}$ quantifies how information is shared between alleles.
For example, in the simple model ($\ref{eq:mtf}$) the kernel is simply
equal to 2 if an allele is compared to itself, 1 otherwise, meaning
that information is uniformly shared across different alleles.
Alternatively, adding supertype-specific features
like~\citet{Heckerman2006Leveraging} would result in a kernel equal to
$3$ between an allele and itself, $2$ between two different alleles
that belong to a common supertype, and $1$ otherwise, resulting in
increased sharing of information within supertypes.

Interestingly this formulation lends itself particularly well to
further generalization. Indeed, for any positive definite kernels
$K_{all}$ and $K_{pep}$ for alleles and peptides, respectively, their
product (\ref{eq:product}) is a valid positive definite kernel over
the product space of pairs
(peptide,allele)~\citep{Aronszajn1950Theory}. This suggests a new
strategy to design predictive models for epitopes across alleles, by
designing specific kernels for alleles and peptides, respectively, and
combining them to learn all allele-specific models simultaneously with
the tensor product kernel (\ref{eq:product}). Benefits of this strategy over the explicit
design and computation of feature vectors $\Phi(x,a)$ are two-folds.
First, it splits the problem of feature vector design into two
subproblems (designing two kernels), each of which can benefit from
previous work on kernel
design~\cite[\emph{e.g.},][]{Schoelkopf2004Kernel}. For example, the
fact that nonlinear kernels such as Gaussian or polynomial kernels for
peptides give good results for SVM trained on individual alleles
suggest that they are natural candidates for the peptide part of the
product kernel. Second, working with kernels alleviates the practical
issues due to the potentially large size of the feature vector
representation $\Phi(x,a)$ in terms of memory for storage or speed of
convergence of algorithms. We now describe in more details the kernels
$K_{pep}$ and $K_{all}$ that can be used for peptides and alleles,
respectively, to create the product kernel used in the application.

\subsection{Peptide kernels}

We consider in this paper mainly peptides made of $9$ amino acids,
although extensions to variable-length peptides poses no difficulty in
principle~\citep{Salomon2006Predicting}. The classical way to represent
these $9$-mers as fixed length vectors is to encode the letter at each
position by a $20$-dimensional binary vector indicating which amino
acid is present, resulting in a $180$-dimensional vector
representations. In terms of kernel, the inner product between two
peptides in this representation is simply the number of letters they
have in common at the same positions, which we take as our baseline
kernel:
\begin{displaymath}
K_{linseq}(x,x') = \sum_{i=1}^{l}\delta(x[i]x'[i]),
\end{displaymath}
where $l$ is the length of the peptides ($9$ in our case), $x[i]$ is
the $i$-th residue in x and $\delta(x[i]x'[i])$ is $1$ if $x[i] =
x'[i]$, $0$ otherwise.

Alternatively, several authors have noted that nonlinear variants of
the linear kernel can improve the performance of SVM for epitope
prediction~\citep{Donnes2002Prediction,Zhao2003Applicationa,Bhasin2004Prediction}. In particular, using a polynomial kernel of degree $p$
over the baseline kernel is equivalent, in terms of feature space, to
encoding $p$-order interactions between amino acids at different
positions. In order to assess the relevance of such non-linear
extensions we tested a polynomial kernel of degree 5, \emph{i.e.},
\begin{displaymath}
  K_{seq5}(x,x') = (K_{linseq}(x,x') +1)^5.
\end{displaymath}

In order to limit the risk of overfitting to the benchmark data we
restrict ourselves to the evaluation of the baseline linear kernel and
its nonlinear polynomial extension. Designing a specific peptide
kernel for epitope prediction, \emph{e.g.}, by weighting differently the
positions known to be critical in the MHC-peptide complex, is however
an interesting research topic that could bring further improvements in
the future.

\subsection{Allele kernels}
Although the question of kernel design for peptides has been raised in
previous studies involving SVM for epitope
prediction~\citep{Donnes2002Prediction,Zhao2003Applicationa,Bhasin2004Prediction,Salomon2006Predicting},
the question of kernel design for alleles is new to our knowledge. We
tested several choices that correspond to previously published
approaches:
\begin{itemize}
\item The \emph{Dirac} kernel is:
\begin{displaymath}
K_{Dirac}(a,a') = \begin{cases}
  1 &\text{ if }a=a'\,,\\
  0&\text{ otherwise.}
\end{cases}
\end{displaymath}
With the Dirac kernel, no information is shared across alleles and the
SVM learns one model for each allele independently from the others.
Therefore this corresponds to the classical setting of learning
epitope prediction models per allele with SVM.
\item The \emph{uniform} kernel is:
  \begin{displaymath}
    K_{uniform}(a,a') = 1\text{ for all }a,a'\,.
  \end{displaymath}
With this kernel all alleles are considered the same, and a unique model is created by pooling together the data available for all alleles.
\item The \emph{multitask} kernel is:
  \begin{displaymath}
K_{multitask} (a,a') = K_{dirac}(a,a') + K_{uniform}(a,a')\,.
  \end{displaymath}
  As explained in the previous section and
  in~\citet{Evgeniou2005Learning} this is the simplest way to train
  different but related models. The SVM learns one model for each
  allele, using known epitopes and non-epitopes for the allele, but
  using also known epitopes and non-epitope for all other alleles with
  a smaller contribution. The training peptides are shared uniformly
  across different alleles.
\item The \emph{supertype} kernel is
  \begin{displaymath}
K_{supertype}(a,a') = K_{multitask} + \delta_{s}(a,a')\,,
  \end{displaymath}
  where $\delta_{s}(a,a')$ is $1$ if $a$ and $a'$ are in the same
  supertype, $0$ otherwise. As explained in the previous section this
  scheme trains a specific models for each allele using training
  peptides from different alleles, but here the training peptides are
  more shared across alleles withing a supertype than across alleles
  in different supertypes. This is used
  by~\citet{Heckerman2006Leveraging}, without the kernel formulation,
  to train a logistic regression model.
\end{itemize}
\citet{Heckerman2006Leveraging} show that the supertype kernel
generally improves the performance of logistic regression models
compared to the uniform or Dirac kernel. Intuitively it seems to be an
interesting way to include prior knowledge about alleles. However, one
should be careful since the definition of supertypes is based on the
comparison of epitopes of different alleles, which suggests that the
supertype information might be based on some information used to
assess the performance of the method in the benchmark experiment. In
order to overcome this issue, and illustrate the possibilities offered
by our formulation, we also tested a kernel between alleles which
tries to quantify the similarity of alleles without using known
epitope information. For that purpose we reasoned that alleles with
similar residues at the positions involved in the peptide binding were
more likely to have similar epitopes, and decided to make a kernel
between alleles based on this information. For each locus we gathered
from~\citet{Doytchinova2004Identifying} the list of positions
involved in the binding site of the peptide (Table \ref{tab:bsite}).
Taking the union of these sets of positions we then represented each
allele by the list of residues at these positions, and used a
polynomial kernel of degree $7$ to compare two lists of residues
associated to two alleles, \emph{i.e},
\begin{displaymath}
  K_{bsite7}(a,a') = \left(\sum_{i\in \textrm{bsite}}\delta(a[i]a'[i])+1\right)^7,
\end{displaymath}
where bsite is the set of residues implied in the binding site for one
of the three allele groups HLA-A, B, C, $a[i]$ is
the $i$-th residue in a and $\delta(a[i]a'[i])$ is $1$ if $a[i] =
a'[i]$, $0$ otherwise.

\subsection{SVM}
We learn epitope models with SVM, a state-of-the-art algorithm for
pattern
recognition~\citep{Vapnik1998Statistical,Scholkopf2002Learning,Schoelkopf2004Kernel}.
We used the \texttt{libsvm} SVM implementation, with a custom kernel
to account for the various kernels we tested, in the PyML environment
(\url{http://pyml.sourceforge.org}).
Besides the kernel, SVM depends on one parameter usually called $C$.
For each experiment, we selected the best $C$ among the values $2^i,
i\in\{-15,-14,\ldots,9,10\}$ by selecting the value leading to the
largest area under the ROC curve estimated by cross-validation on the
training set only. The performance of each method was then tested on
each experiment by evaluating the AUC over the test data.

\section{Data}

In order to evaluate both the performance of our method and the impact
of using various kernels for the peptides or the alleles, we test our
method on three different benchmark datasets that have been compiled
recently to compare the performance of epitope prediction algorithms.

We first use two datasets compiled by \citet{Heckerman2006Leveraging},
where it is already shown that leveraging improves prediction accuracy
with respect to the best published results.The first dataset, called
\textsc{syfpeithy+lanl}, combines experimentally confirmed positive
epitopes from the \textsc{syfpeithy} database~\citep[see][available at
\url{http://www.syfpeithy.de}]{Rammensee1999SYFPEITHI}
and from the Los Alamos HIV database
(\url{http://www.hiv.lanl.gov}) and negative
example randomly drawn from the HLA and amino acid distribution in the
positive examples, for a total of $3152$ data points. For more
details, see~\citet{Heckerman2006Leveraging} where this dataset is
used to compare the leveraged logistic regression with
\emph{DistBoost}. Since this dataset is quite small and was already
used as a benchmark, we use it as a first performance evaluation, and
to compare our kernels.

The second dataset of \citet{Heckerman2006Leveraging} contains
$160,085$ peptides including those from \textsc{sysfpeithy+lanl} and
others from the \textsc{MHCBN} data repository~\citep[see][available
at
\url{http://www.imtech.res.in/raghava/mhcbn/index.html}]{Bhasin2003MHCBN}.
This corresponds to $1,585$ experimentally validated epitopes, and
$158,500$ randomly generated non-binders ($100$ for each positive). We
only kept $50$ negative for each positive in the interest of time and
assuming this would not deteriorate too much the performance of our
algorithm. In the worst case, it is only a handicap for our methods.

Finally, we assess the performance of our method on the MHC-peptide
binding benchmark recently proposed by~\citet{Peters2006community} who
gathered quantitative peptide-binding affinity measurements for
various species, MHC class I alleles and peptide lengths, which makes
it an excellent tool to compare MHC-peptide binding learning methods.
Since our method was first designed for binary classification of HLA
epitopes, we focused on the 9-mer peptides for the $35$ human alleles
and thresholded at $\text{IC}50 = 500$. Nevertheless, the application
of our method to other species or peptide lengths would be
straightforward, and generalization to quantitative prediction should
not be too problematic either. The benchmark contained $29336$
9-mer.

The first dataset is 5-folded, the second 10-folded, so that the test
be only performed on HIV (\textsc{LANL}) data. The third dataset is
5-folded. We used the same folds as~\citet{Heckerman2006Leveraging},
available at
\url{ftp://ftp.research.microsoft.com/users/heckerma/recomb06}
for the first two datasets and the same folds
as~\citet{Peters2006community} available at
\url{http://mhcbindingpredictions.immuneepitope.org/}
for the third one.

Molecule-based allele kernels require the amino-acid sequences
corresponding to each allele. These sequences are available in various
databases,
including\url{http://www.anthonynolan.org.uk/}
and~\citet{Robinson2000IMGT/HLAa}. We used the peptide-sequence
alignment for HLA-A, HLA-B and HLA-C loci. Each sequence was
restricted to residues at positions involved in the binding site of
one of the three loci, see table~\ref{tab:bsite}. Preliminary
experiments showed that using this restriction instead of the whole
sequences didn't change the performance significantly, but it speeds
up the calculation of the kernel. We were not able to find the
sequence of a few molecules of the two datasets
of~\citet{Heckerman2006Leveraging}, so in the experiments implying
these datasets and a molecule-based allele kernel, we used
$K_{bsite7}(a,a') + K_{multitask}(a,a')$ instead of simply using
$K_{bsite7}(a,a')$ , with a sentinel value of $K_{bsite7}(a,a') = 0$
in these cases. This is the sum of two kernels, so still a positive
definite kernel and actually exactly the same thing as $K_{supertype}$
with $K_{bsite7}$ instead of $\delta_s$.

\begin{table}[!t]
\begin{center}
\begin{tabular}{cp{0.8\linewidth}}\hline
Locus & Positions\\\hline
HLA-A & 5, 7, 9, 24, 25, 34, 45, 59, 63, 66, 67, 70, 74, 77, 80, 81, 84, 97, 99, 113, 114, 116, 123, 133, 143, 146, 147, 152, 155, 156, 159, 160, 163, 167, 171\\
HLA-B & 5, 7, 8, 9, 24, 45, 59, 62, 63, 65, 66, 67, 70, 73, 74, 76, 77, 80, 81, 84, 95, 97, 99, 114, 116, 123, 143, 146, 147, 152, 155, 156, 159, 160, 163, 167, 171\\
HLA-C & 5, 7, 9, 22, 59, 62, 64, 66, 67, 69, 70, 73, 74, 77, 80, 81, 84, 95, 97, 99, 116, 123, 124, 143, 146, 147, 156, 159, 163, 164, 167, 171\\\hline
\end{tabular}
\caption{Residue positions involved in the binding site for the three
  loci, according to~\citet{Doytchinova2004Identifying}}
\label{tab:bsite}
\end{center}
\end{table}

\section{Results}

\begin{figure}[!tpb]
  \centerline{\includegraphics[width=.8\linewidth]{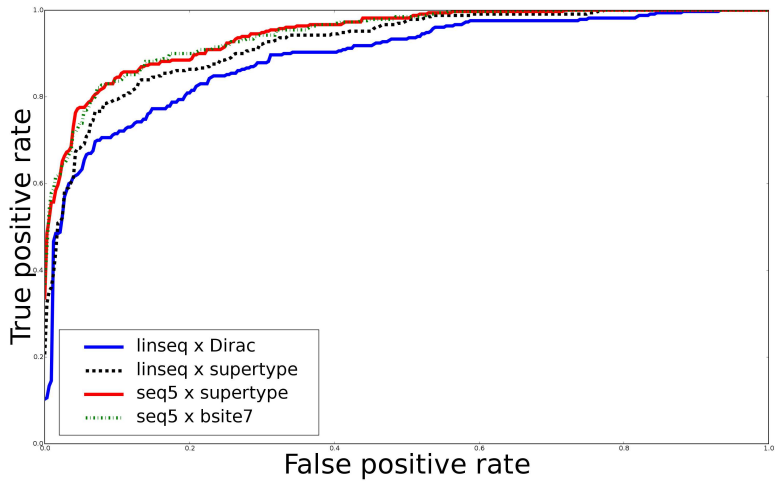}}
  \caption{ROC curves on the pooled five folds of the
    \textsc{syfpeithi+lanl} benchmark.}\label{fig:roc1}
\end{figure}

We first use $K_{linseq}$ and $K_{seq5}$ for the peptides and
$K_{uniform}$ (one SVM for all the alleles), $K_{Dirac}$ (one SVM for
each allele), $K_{multitask}$, $K_{supertype}$ and $K_{bsite7}$ for
the alleles on the small \textsc{syfpeithi+lanl} dataset. Using
combinations of molecule-based and non-molecule-based kernels for
$K_{all}$ didn't improve the prediction, generally the result was as
good as or slightly worse than the result obtained with the best of
the two combined kernels. Results are displayed on
Table~\ref{tab:exp1}, and ROC curves for $K_{linseq}\times K_{Dirac}$,
$K_{linseq}\times K_{supertype}$, $K_{seq5}\times K_{supertype}$ and
$K_{seq5}\times K_{bsite7}$ on figure~\ref{fig:roc1}.

Table~\ref{tab:exp1} demonstrates the benefits of carefully sharing
information across alleles. The \emph{Dirac} allele kernel being the
baseline kernel corresponding to independent training of SVM on
different alleles, we observe an improvement of at least $2\%$ when
information is shared across alleles during training (with the
\emph{multitask,supertype} or \emph{bsite7} strategies). It should be
noted, however, that the \emph{uniform} strategies which amount to
training a single model for all alleles perform considerably worse
than the \emph{Dirac} strategies, justifying the fact that it is still
better to build individual models than a single model for all alleles.
Among the strategies to share information across alleles, the
\emph{supertype} allele kernel seems to work slightly better than the
two other ones. However, one should keep in mind that there is a
possible bias in the performance of the \emph{supertype} kernel,
because some peptides in the test sets might have contributed to the
definition of the allele supertypes. Among the \emph{multitask}
kernel, which considers all different alleles as equally similar, and
the \emph{bsite7} kernel, which shares more information between
alleles that have similar residues at key positions, we observe a
slight benefit for the \emph{bsite7} kernel, which justifies the idea
that including biological knowledge in our framework is simple and
powerful. Finally, we observe that for all allele kernels, the
nonlinear \emph{seq5} peptide kernel outperforms the baseline
\emph{linseq} kernel, confirming that linear models based on
position-specific score matrices might be a too restrictive set of
models to predict accurately epitopes.

In terms of absolute value, all three allele kernels that share
information across alleles combined with the nonlinear \emph{seq5}
peptide kernel (AUC $= 0.943\pm0.015$) strongly outperform the
leveraged logistic regression of~\citet{Heckerman2006Leveraging} (AUC
$=0.906\pm0.016$) and the boosted distance metric learning algorithm
of \citet{Hertz2006PepDist} (AUC $=0.819 \pm 0.055$). This corresponds
to a decrease of roughly $40\%$ of the area above the ROC curve
compared to the best method. As the boosted distance metric learning
approach was shown to be superior to a variety of state-of-the-art
other methods by~\citet{Hertz2006PepDist}, this suggest that our
approach can compete if not overcome the best methods in terms of
accuracy.

As we can clearly see in Table~\ref{tab:exp1}, two factors are
involved in the improvement over the leveraged logistic regression of
\citet{Heckerman2006Leveraging}:
\begin{itemize}
\item The use of an SVM instead of a logistic regression, since this is
  the only difference between the leveraged logistic regression and
  our SVM with a $K_{linseq} \times K_{supertype}$ kernel. This, however, may
  not be intrinsic to the algorithms, but caused by optimization
  issues for the logistic regression in high dimension. 
\item The use of a non-linear kernel for the peptide, as we observe a
  clear improvement in the case of SVM (this improvement might
  therefore also appear if the logistic regression was replaced by a
  kernel logistic regression model with the adequate kernel).
\end{itemize}

Figure~\ref{fig:roc1} illustrates the various improvement underlined
by this experiment: first from the individual SVM ($K_{linseq}\times
K_{Dirac}$), to the $K_{linseq}\times K_{supertype}$ SVM which is the
SVM equivalent of leveraged logistic regression, and finally to
$K_{seq5}\times K_{supertype}$ and $K_{seq5}\times K_{bsite7}$ SVM
that both give better performances than $K_{linseq}\times
K_{supertype}$ SVM because they use a nonlinear kernel to compare the
peptides. It is also worth noting that the \emph{supertype} and the
\emph{bsite7} strategies give very similar results, which makes them
two good strategies to leverage efficiently across the alleles with
different information.

\begin{table}[!t]
\begin{center}
  \begin{tabular}{*{3}{c}}\hline
      $K_{all} \backslash K_{pep}$ & linseq & seq5\\\hline
      uniform & $0.826\pm0.010$ & $0.883\pm0.011$\\
      Dirac  & $0.891\pm0.014$ & $0.893\pm0.024$\\
      multitask & $0.910\pm0.008$ & $0.936\pm0.008$\\
      supertype & $0.923\pm0.011$ & $0.943\pm0.015$\\
      bsite7 & $0.919\pm0.011$ & $0.943\pm0.009$\\\hline
  \end{tabular}
  \caption{AUC results for an SVM trained on the
    \textsc{syfpeithi+lanl} with various kernel and
    estimated error on the 5 folds.}
\label{tab:exp1}
\end{center}
\end{table}

\begin{figure}[!tpb]
  \centerline{\includegraphics[width=.8\linewidth]{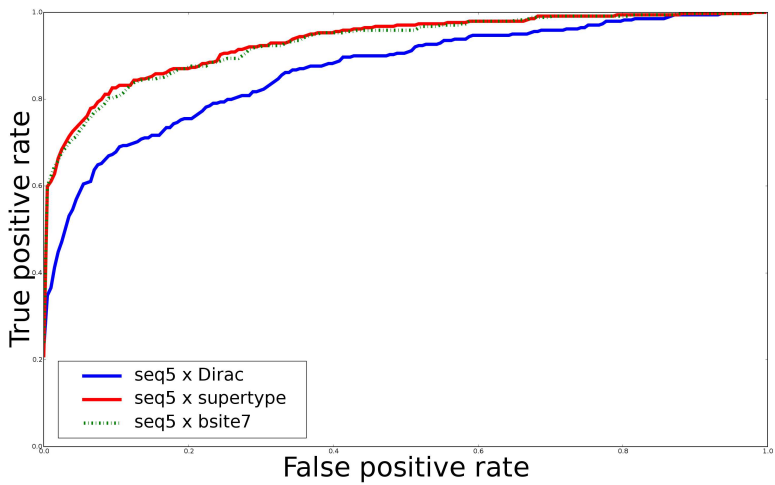}}
  \caption{ROC curves on the pooled ten folds of the
    \textsc{mhcbn+syfpeithi+lanl} benchmark.}\label{fig:roc2}
\end{figure}

These results are confirmed by the \textsc{mhcbn+syfpeithi+lanl}
benchmark, for which the results are displayed in
Table~\ref{tab:exp2}. Again, the use of SVM with our product kernels
clearly improves the performance with respect
to~\citet{Heckerman2006Leveraging} (from $0.906$ to $0.938$).
Moreover, we again observe that learning a leveraged predictor using
the data from all the alleles improves the global performance very
strongly, hence the important step between Dirac ($0.867$) and all the
multitask-based methods, including the simplest multitask kernel
($0.934$). It is worth reminding here that the multitask kernel is
nothing but the sum of the Dirac and uniform kernels, \emph{i.e.},
that it contains no additional biological information: the improvement
is caused by the mere fact of using roughly (with a pondering of
$0.5$) the points of other alleles to learn the predictor of one
allele. Figure~\ref{fig:roc2} show the ROC curves for SVM with
$K_{seq5}\times K_{Dirac}$, $K_{seq5}\times K_{supertype}$ and
$K_{seq5}\times K_{bsite7}$ kernels on this benchmark. Again, we
clearly see the strong improvement between leveraged and non-leveraged
strategies. The difference between the $K_{seq5}\times K_{Dirac}$ and
the two others is only caused by leveraging, since in the three case
the same nonlinear strategy was used for the peptide part. On the
other hand, the figure illustrates once again that our two high-level
(\emph{i.e.}, more sophisticated than \emph{multitask}) strategies for
leveraging across alleles give almost the same result.

\begin{table}[!t]
\begin{center}
  \begin{tabular}{l c}\hline
      Method & AUC\\\hline
      Leveraged LR & $0.906$\\
      $K_{linseq}\times K_{stype}$ & $0.916\pm0.008$\\
      $K_{seq5}\times K_{dirac}$ & $0.867\pm0.010$\\
      $K_{seq5}\times K_{multitask}$ & $0.934\pm0.006$\\
      $K_{seq5}\times K_{stype}$ & $0.939\pm0.006$\\
      $K_{seq5}\times K_{bsite7}$ & $0.938\pm0.006$\\\hline
  \end{tabular}
  \caption{AUC results for an SVM trained on the
    \textsc{mhcbn+syfpeithi+lanl} benchmark with various kernel and
    estimated error on the 10 folds.}
  \label{tab:exp2}
\end{center}
\end{table}

Finally, Table~\ref{tab:peters} presents the performance on the
\textsc{iedb} benchmark proposed in~\citet{Peters2006community}. The
indicated performance corresponds, for each method, to the average on
the AUC for each of the $35$ alleles. This gives an indication of the
global performances of each methods. The ANN field is the tool
proposed in~\citet{Peters2006community} giving the best results on the
9-mer dataset, an artificial neural network proposed
in~\citet{Nielsen2003Reliable}, while the ADT field refers to the
adaptive double threading approach recently proposed
in~\cite{Jojic2006Learning} and tested on the same benchmark.  These
tools were compared to and significantly outperformed other tools in
the comprehensive study of~\citet{Peters2006community},
specifically~\citet{Peters2005Generating}
and~\citet{Bui2005Automated}, that are both scoring-matrix-based. Our
approach gives equivalent results in terms of global performances
as~\citet{Nielsen2003Reliable}, and therefore outperforms the other
internal methods.

Table~\ref{tab:peters-small} presents the performances on the $10$
alleles with less than $200$ training points, together with the
performances of the best internal tool, \citet{Nielsen2003Reliable}
ANN, and the adaptive double threading model that gave good prediction
performances on the alleles with few training data. Except for one
case, our SVM outperforms both models. This means of course that our
approach does not perform as well as~\citet{Nielsen2003Reliable} on
the alleles with a large training set, but nothing prevents an
immunologist from using one tool for some alleles and another tool for
other alleles. As we said in introduction, our original concern was to
improve binding prediction for alleles with few training points, and
for which it is hard to generalize. This was the main point of using a
multitask learning approach. The results on this last benchmark
suggest that the leveraging approaches succeed in improving prediction
performances when few training points are available.

\begin{table}[!t]
\begin{center}
  \begin{tabular}{l c c}\hline
      Method & AUC\\\hline
      SVM with $K_{seq5}\times K_{Dirac}$& $0.804$\\
      SVM with $K_{seq5}\times K_{supertype}$&  $0.877$\\
      SVM with $K_{seq5}\times K_{bsite7}$& $0.892$\\
      ADT & $0.874$\\
      ANN & $0.897$\\\hline
  \end{tabular}
  \caption{AUC results for an SVM trained on the
    \textsc{iedb} benchmark with various methods.}
\label{tab:peters}
\end{center}
\end{table}

\begin{table}[!t]
\begin{center}
  \begin{tabular}{*{5}{c}}\hline
Allele & Peptide number & $K_{seq5}\times K_{bsite7}$ & ADT & ANN \\\hline
A\_2301 & $104$ & $0.887 \pm 0.021$& $0.804$ & $0.852$ \\
A\_2402 & $197$ & $0.826 \pm 0.025$& $0.785$ & $0.825$ \\
A\_2902 & $160$ & $0.948 \pm 0.015$& $0.887$ & $0.935$ \\
A\_3002 & $92$  & $0.826 \pm 0.048$& $0.763$ & $0.744$ \\
B\_1801 & $118$ & $0.866 \pm 0.020$& $0.869$ & $0.838$ \\
B\_4002 & $118$ & $0.796 \pm 0.025$& $0.819$ & $0.754$ \\
B\_4402 & $119$ & $0.782 \pm 0.084$& $0.678$ & $0.778$ \\
B\_4403 & $119$ & $0.796 \pm 0.042$& $0.624$ & $0.763$ \\
B\_4501 & $114$ & $0.889 \pm 0.029$& $0.801$ & $0.862$ \\
B\_5701 & $59$ & $0.938 \pm 0.046$& $0.832$ & $0.926$ \\\hline
\end{tabular}
  \caption{Detail of the \textsc{iedb} benchmark for the $10$ alleles
    with less than $200$ training points (9-mer data).}
\label{tab:peters-small}
\end{center}
\end{table}

\section{Discussion and concluding remarks}

In this paper, we introduced a general framework to share efficiently
the binding information available for various alleles by simply
defining a kernel for the peptides, and another one for the alleles.
The result is a simple model for MHC-peptide binding prediction that
uses information from the whole dataset to make specific prediction
for any of the alleles. Our approach is simple, general and both easy
to adapt to a specific problem by using more adequate kernels, and to
implement, by running any SVM implementation with these kernels.
Everything is performed in low dimension and with no need for feature
selection.

We presented performances on three benchmarks. On the first two
benchmark, our approach performed considerably better than the
state-of-the-art, which illustrates the good general behavior in terms
of prediction accuracy. Besides, these experiments clearly confirmed
the interest of leveraging the information across the alleles. On the
last benchmark, the results were globally comparable to the best
state-of-the-art tested in~\citet{Peters2006community}, with a strong
improvement on the alleles for which few training points were
available, probably, as it was already observed, because of the fact
that our model uses all the points from all the alleles for each
allele-specific prediction.

Another contribution is the use of allele sequences, which allows us
to improve the prediction accuracy and to do as well as what was done
with the supertype information. Supertype is a crucial information and
a key concept in the development of epitope-based vaccines, for
example to find epitopes that bind several alleles instead of just
one. However, one should be careful when using it to learn an
automatic epitope predictor because even if the idea behind a
supertype definition is to represent a general ligand trend, the
intuition is always guided by the fact that some alleles have
overlapping repertoires of known binders, and it is not easy to figure
out to which extent the known epitopes used to assess the predictor
performances were used to design the supertypes.

Because of these overfitting issues and the fact that supertypes are
difficult to define, the good performances of molecule-based allele
kernel with respect to the supertype-based allele kernels are good
news. This potentially allows us to leverage efficiently across
alleles even when the supertype is unknown, which is often the case,
and we don't take the risk to use overfitted information when learning
on large epitope databases.

Although the kernels we used already gave good performances, there is
still room for improvement. A first way to improve the performances
would be to use more adequate kernels to compare the peptides and,
probably more important, to compare the alleles. In other words
answering the question, what does it mean in the context of
MHC-peptide binding prediction for two alleles to be similar? Possible
answers should probably involve better kernels for the allele
sequences, and structural information which could be crucial to
predict binding and, as we said in introduction, is already used in
some models. Another interesting possibility is, as it was suggested
in~\citet{Hertz2007Identifying}, the use of true non-binders, that
could make the predictor more accurate than randomly generated
peptides since these experimentally assessed peptides are in general
close to the known binders. Finally, it could be useful to incorporate
the quantitative IC50 information when available, instead of simply
thresholding as we did for the last benchmark.

This leads us to the possible generalizations we hope to work on,
besides these improvements. Using the binding affinity information, it
is obviously possible to apply our general framework to predict
quantitative values, using regression models with the same type of
kernels. This framework could also be used for a lot of similar
problems involving binding, like MHC-type-II-peptide binding where
sequences can have variable length and the alignment of epitopes
usually performed as pre-processing can be ambiguous.
\citet{Salomon2006Predicting} already proposed a kernel for this case.
Another interesting application would be drug design, for example
protein-kinase-inhibitor binding prediction, or prediction of a virus
susceptibility to a panel of drugs for various mutations of the virus.

\bibliographystyle{natbib}

\end{document}